\documentclass[3p,times,procedia]{elsarticle}
\usepackage{nupha_ecrc}

\volume{00}
\firstpage{1}

\journalname{Nuclear Physics A}

\runauth{}

\jid{nupha}

\jnltitlelogo{Nuclear Physics A}

\usepackage{amssymb}
\usepackage{lineno}
\usepackage[figuresright]{rotating}

\newcommand{\pp}{{\rm pp}}
\newcommand{\pbpb}{{\rm{Pb-Pb}}}
\newcommand{\auau}{{\rm{Au-Au}}}

\newcommand{\jpsi}{{\rm{J}/\psi}}
\newcommand{\psiprime}{{\psi{\rm(2S)}}}
\newcommand{\snn}{{\sqrt{s_{\rm NN}}}}
\newcommand{\raa}{{R_{\rm AA}}}
\newcommand{\smallraa}{{r_{\rm AA}}}
\newcommand{\avrg}[1] {\langle #1\rangle}
\newcommand{\pt}{{p_{\rm T}}}

\newcommand{\npart}{{N_{\rm part}}}
\newcommand{\tab}{{T_{\rm AB}}}
\newcommand{\meanptsquare}{{\avrg{\pt^2}}}

\begin{document}
\begin{frontmatter}
\dochead{}
\title{Charmonium production in $\pbpb$ collisions with ALICE at the LHC}
\author[IRFU]{H. Pereira Da Costa}
\author{for the ALICE Collaboration}
\address[IRFU]{Commissariat \`a l’Energie Atomique, IRFU, Saclay, France}

\begin{abstract}
We report on published charmonium measurements performed by ALICE, at the LHC, in $\pbpb$ collisions at a center of mass energy per nucleon-nucleon collision $\snn = 2.76$ TeV, at both mid ($|y|<0.8$) and forward ($2.5<y<4$) rapidities. 
The nuclear modification factor of inclusive $\jpsi$ is presented as a function of the collision centrality and the $\jpsi$ transverse momentum, $\pt$.
The variation of the $\jpsi$ mean transverse momentum square as a function of the collision centrality is also discussed. 
These measurements are compared to state of the art models that include one or several of the following mechanisms: color screening of the charm quarks, statistical hadronization at the QGP phase boundary, balance between $\jpsi$ dissociation and regeneration in the QGP, $\jpsi$ interaction with a dense comoving medium. 
Results on the production of the heavier and less bound $\psiprime$ meson in $\pbpb$ collisions at forward-rapidity are also presented and compared to both models and measurements performed by other experiments. At mid-rapidity we also report on ALICE unique capability to separate prompt and non-prompt $\jpsi$ production down to low $\pt$ ($\geq 1.5$ GeV/$c$) and thus disentangle between effects on prompt $\jpsi$ mesons and energy loss of $b$ quarks in the QGP. 
\end{abstract}


\end{frontmatter}


\vspace*{-3mm}
\section*{}
\label{}

Charmonia (for instance $\jpsi$ and $\psiprime$) are mesons formed of a charm and anti-charm quark pair. 
In relativistic Heavy-Ion (HI) collisions, such as those delivered by the LHC in 2010 and 2011, their production is expected to be altered by the formation of a Quark-Gluon Plasma (QGP), a state of the nuclear matter predicted by Lattice QCD calculations at high temperature and energy density and for which quarks and gluons are deconfined. Such modifications are quantified using the nuclear modification factor $\raa$, which is the ratio between the charmonium yield measured in HI collisions and the cross section measured in $\pp$ at the same energy, further normalized by the nuclear overlap function $\tab$. A suppression of the charmonium production in HI collisions with respect to $\pp$ corresponds to $\raa<1$ whereas an enhancement corresponds to $\raa>1$.

\begin{figure}[h]
\begin{center}
\begin{tabular}{cc}
\includegraphics[width=0.48\linewidth,keepaspectratio]{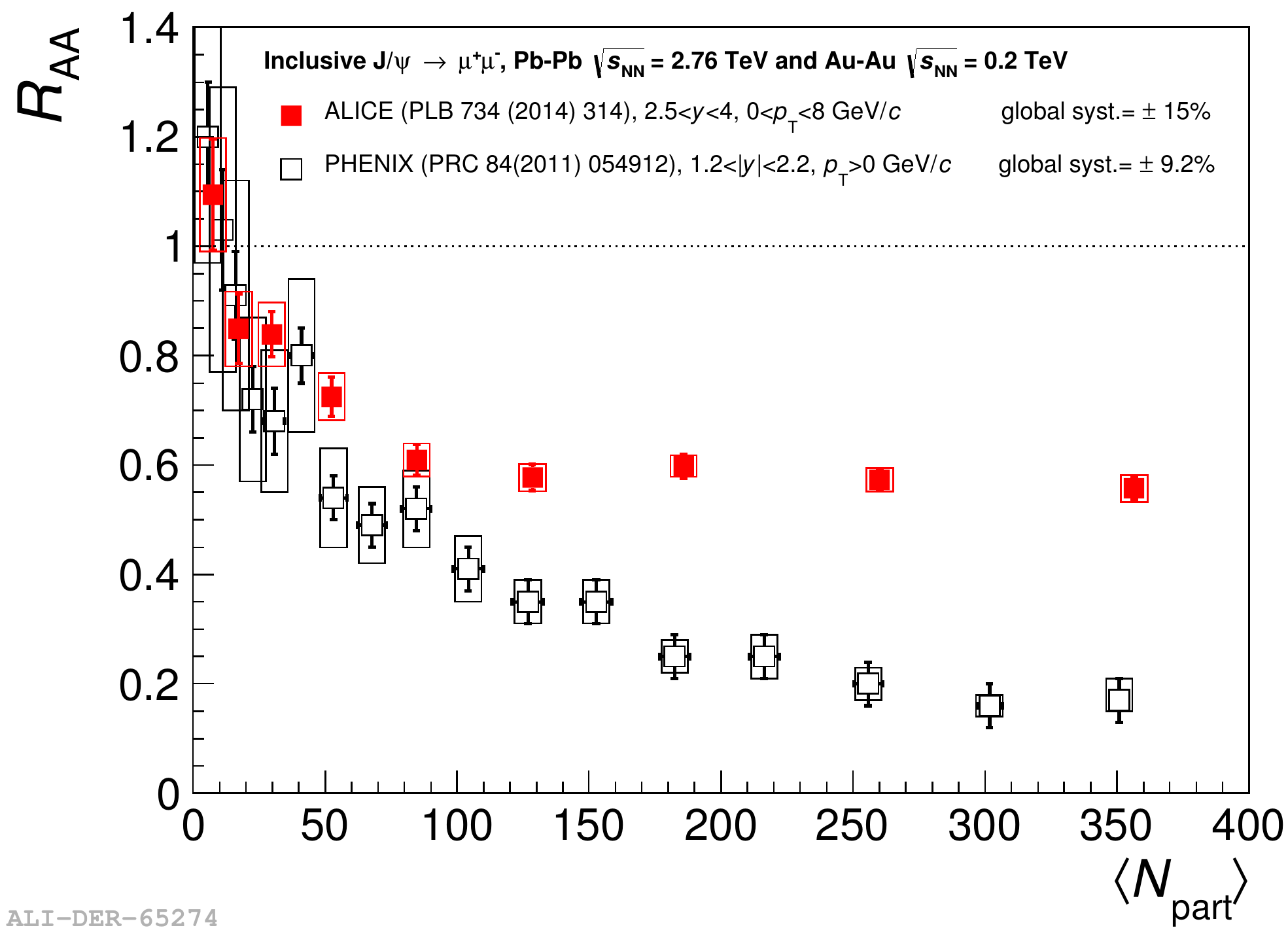}&
\includegraphics[width=0.48\linewidth,keepaspectratio]{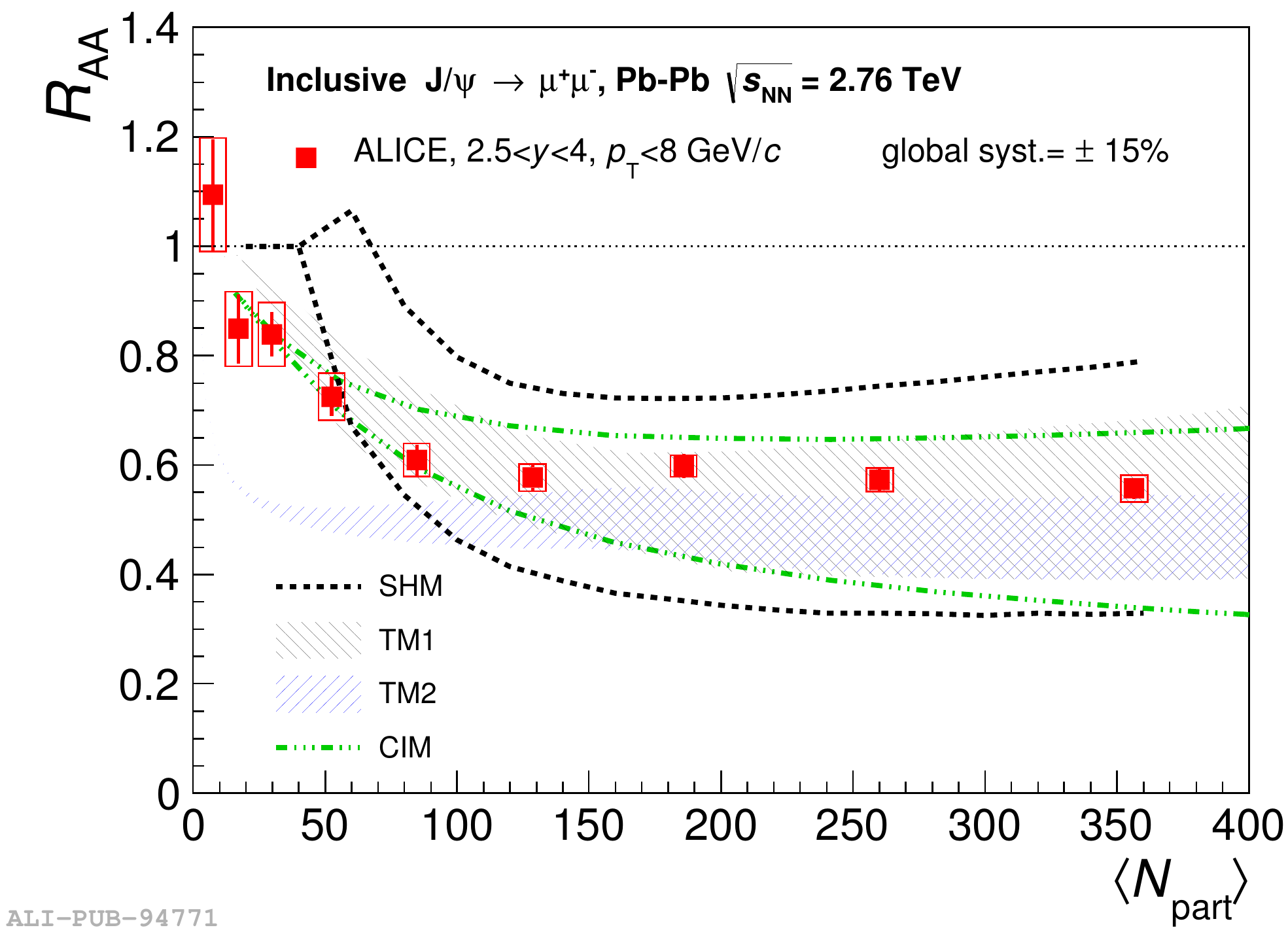}
\end{tabular}
\end{center}
\vspace*{-3mm}
\caption{\label{raa_forward_integrated} $\jpsi$ $\raa$ at forward rapidity measured by ALICE in $\pbpb$ collisions at $\snn=2.76$~TeV as a function of $\npart$ compared to PHENIX measurement (left) and theoretical models (right).}
\end{figure} 

The left panel of Fig.~\ref{raa_forward_integrated} shows the $\jpsi$ $\raa$ measured by ALICE at forward rapidity ($2.5<y<4$) in $\pbpb$ collisions at $\snn = 2.76$ TeV as a function of the number of nucleons participating to the collision $\npart$~\cite{Abelev:2013ila}. It is compared to a measurement by PHENIX in $\auau$ collisions at $\snn=0.2$~TeV~\cite{Adare:2011yf}. In both cases a suppression is observed for central collisions (large values of $\npart$), attributed for the most part to the formation of the QGP. It is however less pronounced for ALICE than PHENIX, although the collision energy is more than 10 times larger. This property has been attributed to the onset of a recombination component to the $\jpsi$ production at LHC energies, which counterbalances the suppression observed already at lower energies. This is confirmed by several model calculations shown in the right panel of Fig.~\ref{raa_forward_integrated} and which are all able to reproduce ALICE measurements for central collisions, provided that a charm quark recombination component is added~\cite{Andronic:2011yq,Zhao:2011cv,Zhou:2014kka,Ferreiro:2012rq}. Similar conclusions can be reached from mid-rapidity measurements, albeit with larger uncertainties~\cite{Adam:2015isa}. 

Details on the mechanism by which $\jpsi$ mesons are suppressed and (re)generated depend on the model and more differential comparisons are needed in order to discriminate between them. One can for instance consider the $\pt$ dependence of the $\jpsi$ $\raa$ for different bins of centrality or, the other way around, the $\jpsi$ $\raa$ as a function of $\npart$ for different bins of $\pt$, as discussed in~\cite{Adam:2015isa}. For the moment, most of the models shown in Fig.~\ref{raa_forward_integrated} are capable of reproducing these differential results too, at least qualitatively, but more systematic comparisons should be performed. In particular, for peripheral collisions extra attention must be given to the fact that at very low $\pt$ (typically $\pt<300$~MeV/$c$), a fraction of the measured $\jpsi$ might proceed from coherent photo-production instead of hadro-production and is not expected to scale with $\tab$ in absence of nuclear effects~\cite{Adam:2015gba}. Such $\jpsi$s should be properly accounted for or even removed from the measurement before comparing to the models cited above.

\begin{figure}[h]
\begin{center}
\includegraphics[width=0.48\linewidth,keepaspectratio]{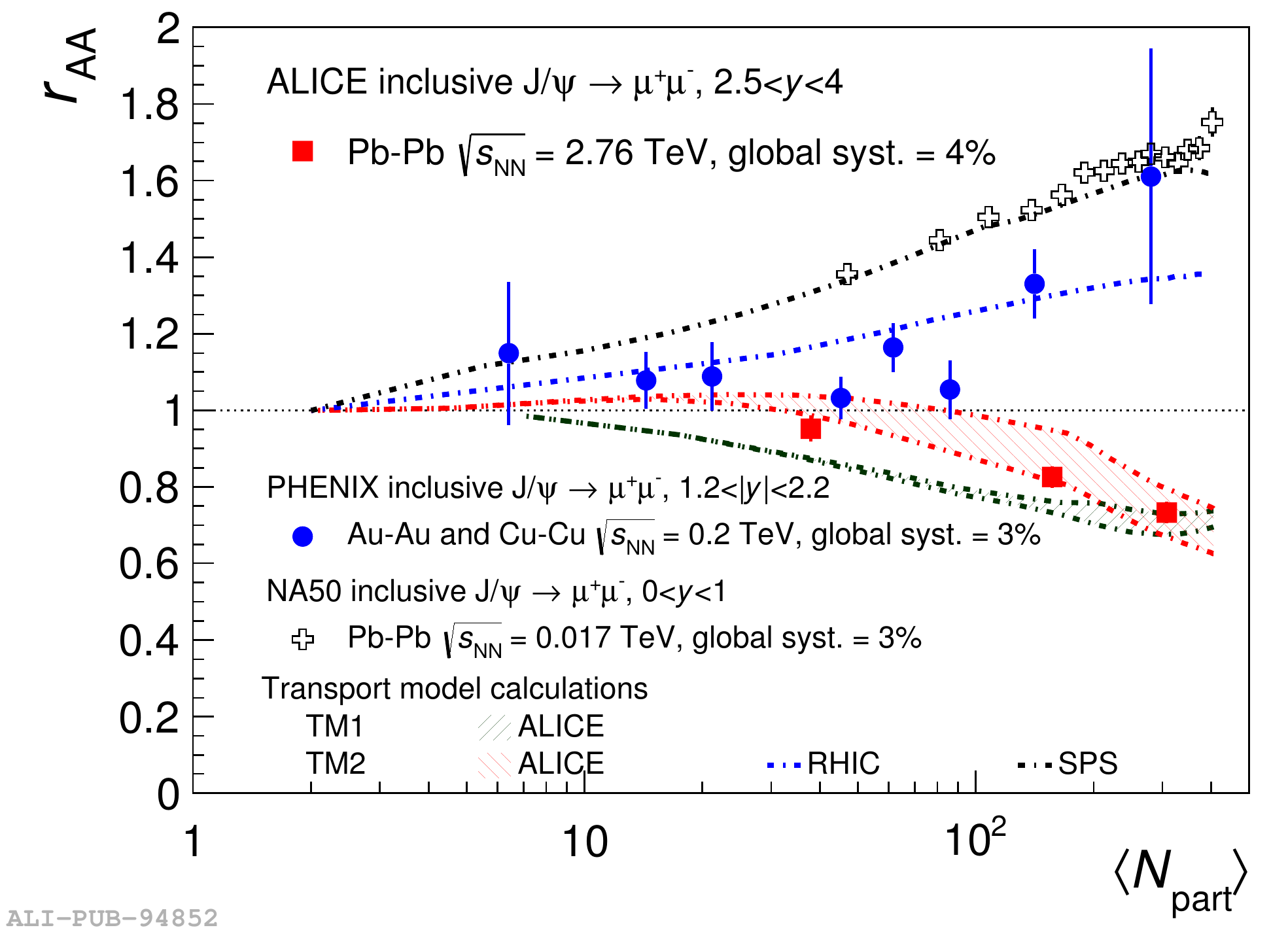}
\end{center}
\vspace*{-3mm}
\caption{\label{meanpt} Ratio $\smallraa$ between $\jpsi$ $\meanptsquare$ in HI and in pp collisions as a function of $\npart$ measured by NA50, PHENIX and ALICE. Measurements are compared to two transport models referred to as TM1~\cite{Zhao:2011cv} and TM2~\cite{Zhou:2014kka}.}
\end{figure} 

Measuring the $\jpsi$ mean transverse momentum square $\meanptsquare$ as a function of the colliding species and the collision centrality provides another mean to study medium effects on the $\jpsi$ production. In order to quantify differences between HI collisions and $\pp$, the ratio $\smallraa = \meanptsquare_{\rm AA}/\meanptsquare_{pp}$ is formed. Measurements of $\smallraa$ as a function of $\npart$ for NA50~\cite{Abreu:2000xe}, PHENIX~\cite{Adare:2011yf,Adare:2008sh} and ALICE~\cite{Adam:2015isa} at forward rapidity are shown in Fig.~\ref{meanpt}. The variation of $\smallraa$ with respect to $\npart$ exhibits a striking energy dependence. At low energy ($\snn=0.017$~TeV), $\smallraa$ increases with increasing $\npart$, which is attributed to the Cronin effect~\cite{PhysRevD.11.3105}. On the contrary, $\smallraa$ shows no strong dependence on $\npart$ at intermediate energy ($\snn = 0.2$~TeV) and decreases with increasing $\npart$ at LHC energy. This progressive change of trend is attributed to the onset of $\jpsi$ produced by recombination in the plasma, and can be reproduced consistently by the transport model from~\cite{Zhou:2014kka}. As was the case for $\raa$, the uncertainties on the mid-rapidity $\smallraa$ measurements performed by ALICE are larger than at forward rapidity. The values are however significantly below the ones measured at lower energy and no visible centrality dependence is observed, which is not well reproduced by models~\cite{Adam:2015rba}.

\begin{figure}[h]
\begin{center}
\begin{tabular}{cc}
\includegraphics[width=0.48\linewidth,keepaspectratio]{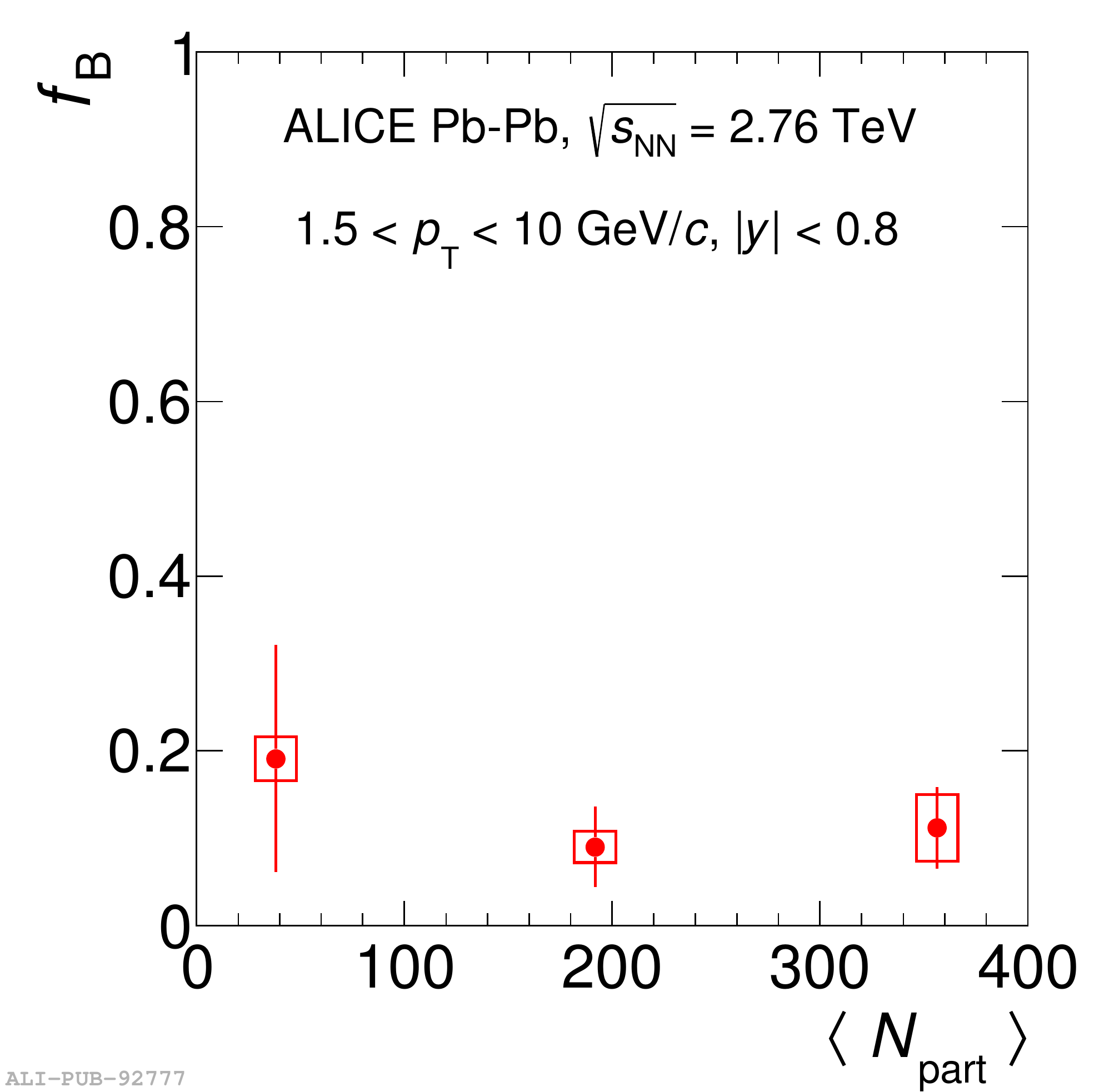}&
\includegraphics[width=0.48\linewidth,keepaspectratio]{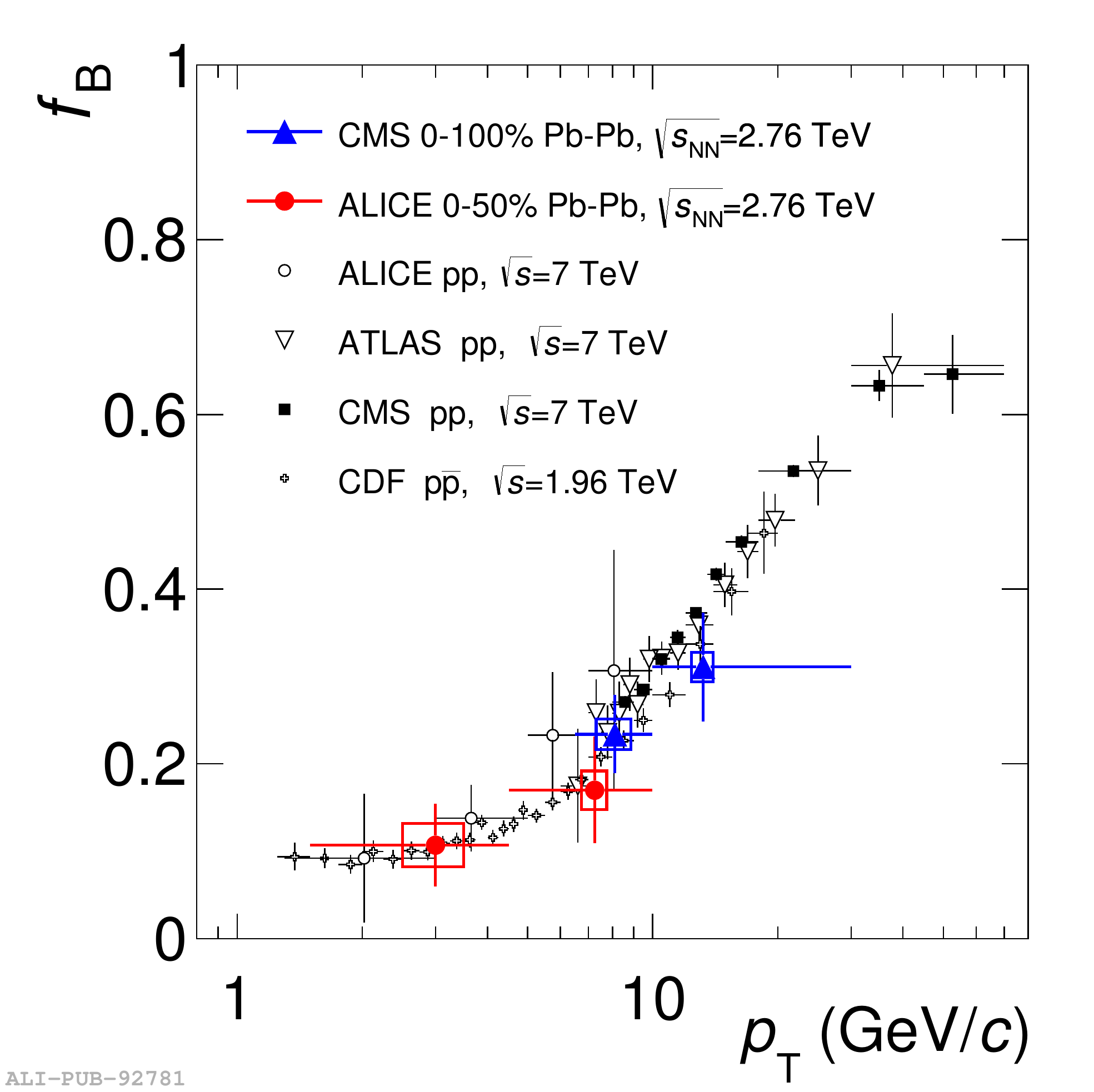}
\end{tabular}
\end{center}
\vspace*{-3mm}
\caption{\label{fb} Ratio of non-prompt to inclusive $\jpsi$ yields measured at mid-rapidity ($|y|<0.8$), as a function of $\npart$ (left) and $\pt$ (right). Results are compared to similar measurements by CMS in $\pbpb$ collisions as well as by ALICE, ATLAS, CMS and CDF in $\pp$ collisions~\cite{Adam:2015rba}.}
\end{figure} 

The high-resolution vertexing capabilities provided by the ALICE Inner Tracking System~\cite{Aamodt:2010aa} allow to separate prompt and non-prompt $\jpsi$ production at mid-rapidity in both $\pp$ and HI collisions. Prompt $\jpsi$s correspond to direct production and decays from higher mass excited charmonium states, whereas non-prompt $\jpsi$s originate from $b$-hadron decays and provide a measurement of $b$-quark energy loss in the QGP. The fraction of non-prompt to inclusive $\jpsi$ yields, $f_b$, is shown in Fig.~\ref{fb} as a function of $\npart$ for $1.5<\pt<10$~GeV/$c$ (left) and as a function of $\pt$ for $0-50$~\% collision centralities (right)~\cite{Adam:2015rba}. No significant dependence on centrality is observed. Moreover, values of $f_b$ measured in $\pp$ and $\pbpb$ collisions do not differ significantly within uncertainties. Consequently, the inclusive and prompt $\jpsi$ $\raa$ are very close and all the observations performed on the inclusive $\jpsi$ $\raa$ also apply to prompt $\jpsi$s, at least at mid-rapidity.

\begin{figure}[h]
\begin{center}
\begin{tabular}{cc}
\includegraphics[width=0.48\linewidth,keepaspectratio]{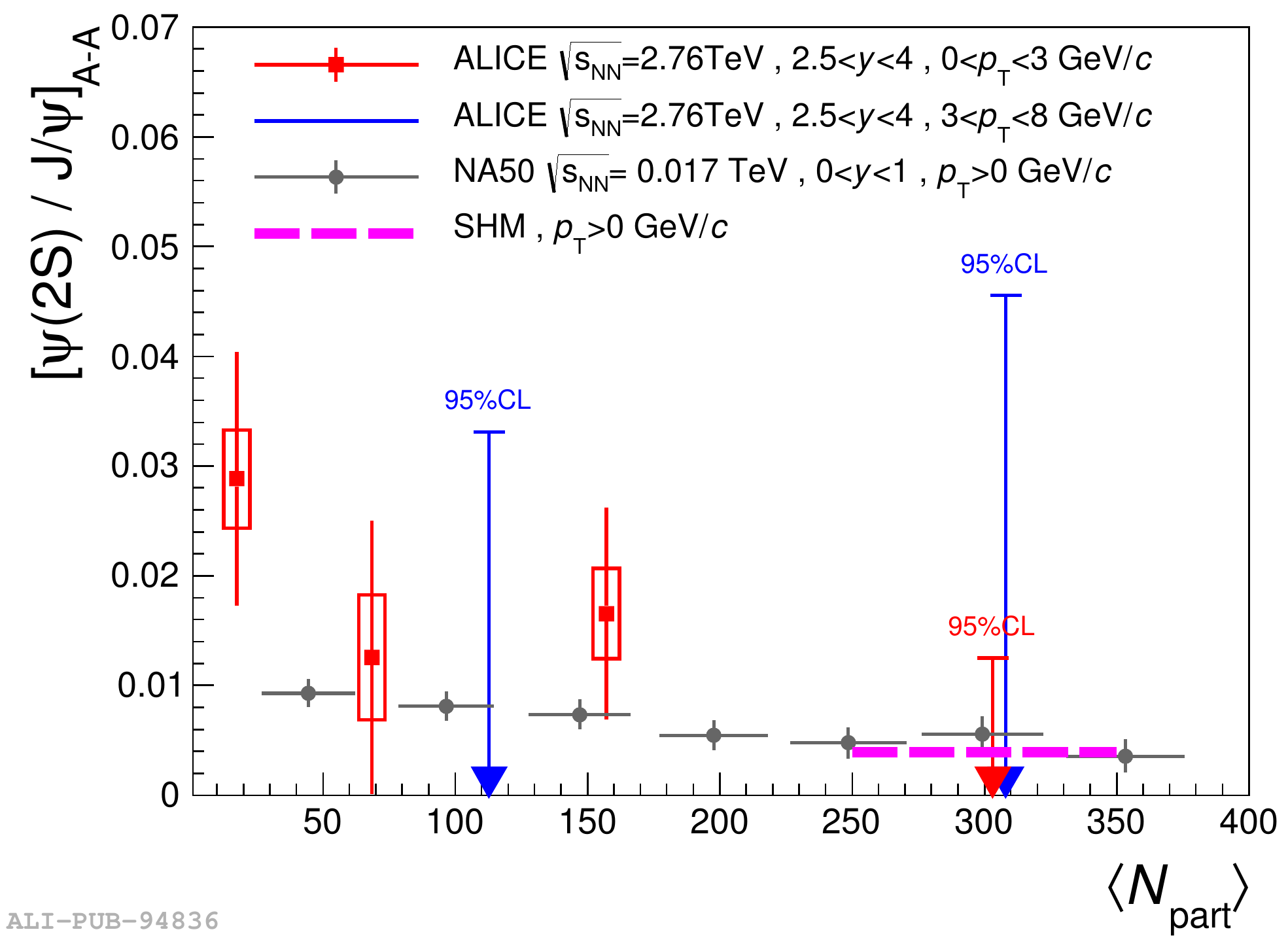}&
\includegraphics[width=0.48\linewidth,keepaspectratio]{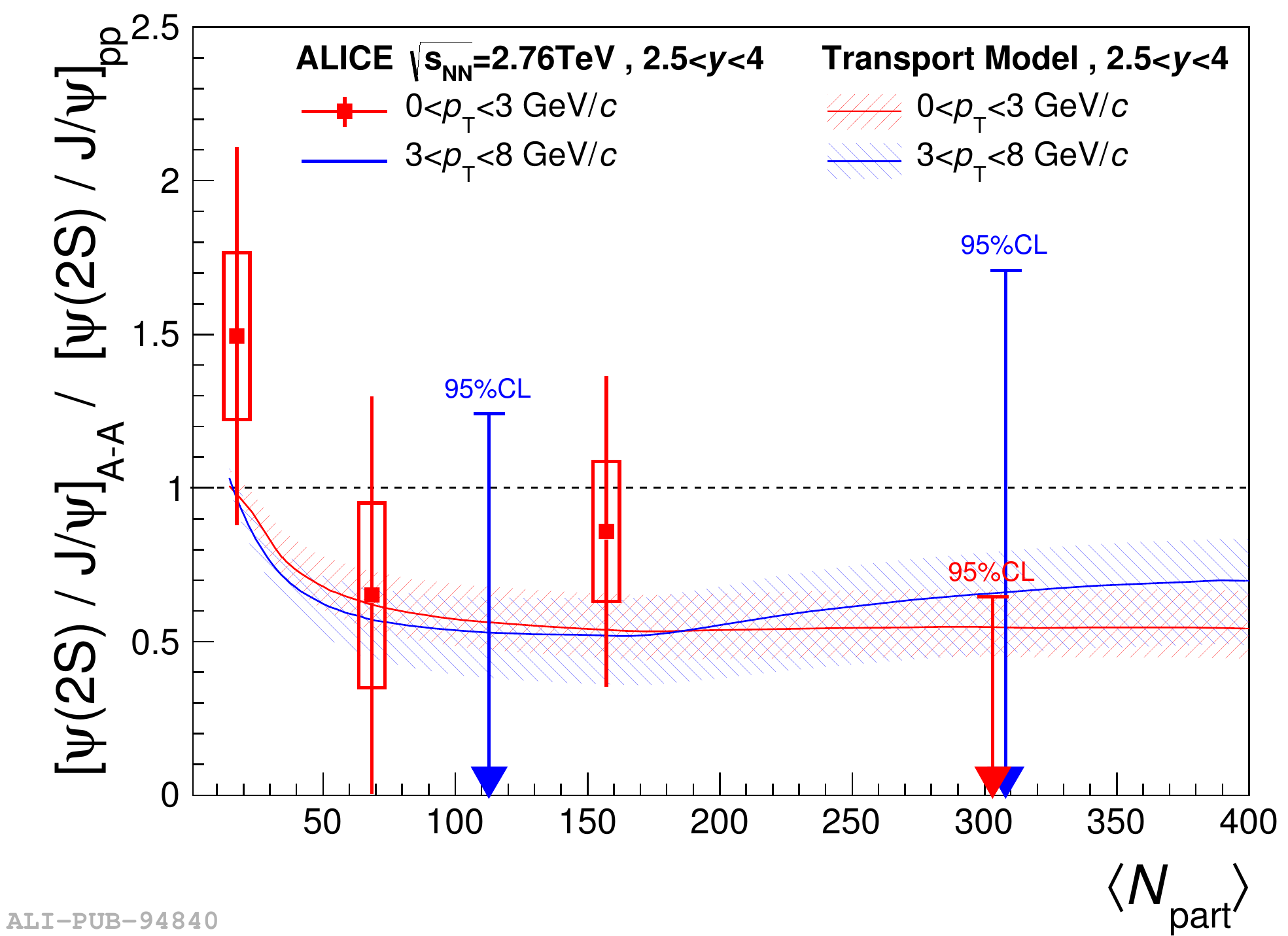}
\end{tabular}
\end{center}
\vspace*{-3mm}
\caption{\label{psiprime} Left: $\psiprime$-to-$\jpsi$ ratio as a function of $\npart$ for two $\pt$ intervals, compared to measurements from NA50~\cite{Alessandro:2006ju} and the statistical hadronization model~\cite{Andronic:2009sv}. Right: $\pbpb$-to-$\pp$ $\psiprime$-to-$\jpsi$ double ratio as a function of $\npart$ compared to a transport model~\cite{Chen:2013wmr}.}
\end{figure} 

Concerning the production of $\psiprime$ mesons in HI collisions, ALICE has measured the $\psiprime$-to-$\jpsi$ ratio in $\pbpb$ at $\snn=2.76$~TeV as well as the $\pbpb$ over $\pp$ $\psiprime$-to-$\jpsi$ double ratio as a function of $\npart$~\cite{Adam:2015isa}. These ratios could be used to discriminate between the various mechanisms invoked for the recombination component of charmonium production in HI collisions. Results are presented in Fig.~\ref{psiprime}. For the moment, statistics is rather limited and for most central collisions, only 95~\% confidence levels are measured. The measurements are well reproduced by both the statistical hadronization model~\cite{Andronic:2009sv} and a transport model~\cite{Chen:2013wmr}. Note however that there exists some tension between these results and the one reported by CMS in a slightly different rapidity range~\cite{Khachatryan:2014bva}. 

To summarize: (i) measurements of the $\jpsi$ $\raa$ (either integrated or differential) as well as its $\meanptsquare$ are consistent with the presence of a regeneration component in the measured production yield, although no consensus exist on the mechanism of this regeneration; (ii) the contribution of non-prompt $\jpsi$ from $b$-hadron decays to the inclusive yields has no significant impact on the measured inclusive $\raa$ and (iii) medium effects on $\psiprime$ production in HI collisions are still largely unconstrained and some tension exists among available measurements. New data from LHC Run2 will certainly help with that matter.

\bibliographystyle{elsarticle-num}
\bibliography{Hugo_Pereira_proceedings_QM2015.bib}

\begin{thebibliography}{10}
\expandafter\ifx\csname url\endcsname\relax
  \def\url#1{\texttt{#1}}\fi
\expandafter\ifx\csname urlprefix\endcsname\relax\def\urlprefix{URL }\fi
\expandafter\ifx\csname href\endcsname\relax
  \def\href#1#2{#2} \def\path#1{#1}\fi

\bibitem{Abelev:2013ila}
B.~B. Abelev, et~al., {Centrality, rapidity and transverse momentum dependence
  of $J/\psi$ suppression in Pb-Pb collisions at $\sqrt{s_{\rm NN}}$=2.76 TeV},
  Phys. Lett. B734 (2014) 314--327.
\newblock \href {http://arxiv.org/abs/1311.0214} {\path{arXiv:1311.0214}},
  \href {http://dx.doi.org/10.1016/j.physletb.2014.05.064}
  {\path{doi:10.1016/j.physletb.2014.05.064}}.

\bibitem{Adare:2011yf}
A.~Adare, et~al., {$J/\psi$ suppression at forward rapidity in Au+Au collisions
  at $\sqrt{s_{NN}}=200$ GeV}, Phys. Rev. C84 (2011) 054912.
\newblock \href {http://arxiv.org/abs/1103.6269} {\path{arXiv:1103.6269}},
  \href {http://dx.doi.org/10.1103/PhysRevC.84.054912}
  {\path{doi:10.1103/PhysRevC.84.054912}}.

\bibitem{Andronic:2011yq}
A.~Andronic, P.~Braun-Munzinger, K.~Redlich, J.~Stachel, {The thermal model on
  the verge of the ultimate test: particle production in Pb-Pb collisions at
  the LHC}, J. Phys. G38 (2011) 124081.
\newblock \href {http://arxiv.org/abs/1106.6321} {\path{arXiv:1106.6321}},
  \href {http://dx.doi.org/10.1088/0954-3899/38/12/124081}
  {\path{doi:10.1088/0954-3899/38/12/124081}}.

\bibitem{Zhao:2011cv}
X.~Zhao, R.~Rapp, {Medium Modifications and Production of Charmonia at LHC},
  Nucl. Phys. A859 (2011) 114--125.
\newblock \href {http://arxiv.org/abs/1102.2194} {\path{arXiv:1102.2194}},
  \href {http://dx.doi.org/10.1016/j.nuclphysa.2011.05.001}
  {\path{doi:10.1016/j.nuclphysa.2011.05.001}}.

\bibitem{Zhou:2014kka}
K.~Zhou, N.~Xu, Z.~Xu, P.~Zhuang, {Medium effects on charmonium production at
  ultrarelativistic energies available at the CERN Large Hadron Collider},
  Phys. Rev. C89~(5) (2014) 054911.
\newblock \href {http://arxiv.org/abs/1401.5845} {\path{arXiv:1401.5845}},
  \href {http://dx.doi.org/10.1103/PhysRevC.89.054911}
  {\path{doi:10.1103/PhysRevC.89.054911}}.

\bibitem{Ferreiro:2012rq}
E.~G. Ferreiro, {Charmonium dissociation and recombination at LHC: Revisiting
  comovers}, Phys. Lett. B731 (2014) 57--63.
\newblock \href {http://arxiv.org/abs/1210.3209} {\path{arXiv:1210.3209}},
  \href {http://dx.doi.org/10.1016/j.physletb.2014.02.011}
  {\path{doi:10.1016/j.physletb.2014.02.011}}.

\bibitem{Adam:2015isa}
J.~Adam, et~al., {Differential studies of inclusive J/$\psi$ and $\psi$(2S)
  production at forward rapidity in Pb-Pb collisions at $\mathbf{\sqrt{{\textit
  s}_{_{NN}}}}$ = 2.76 TeV}\href {http://arxiv.org/abs/1506.08804}
  {\path{arXiv:1506.08804}}.

\bibitem{Adam:2015gba}
J.~Adam, et~al., {Measurement of an excess in the yield of J/$\psi$ at very low
  $p_{\rm T}$ in Pb-Pb collisions at $\sqrt{s_{\rm NN}}$ = 2.76 TeV}\href
  {http://arxiv.org/abs/1509.08802} {\path{arXiv:1509.08802}}.

\bibitem{Abreu:2000xe}
M.~C. Abreu, et~al., {Transverse momentum distributions of J / psi, psi-prime,
  Drell-Yan and continuum dimuons produced in Pb Pb interactions at the SPS},
  Phys. Lett. B499 (2001) 85--96.
\newblock \href {http://dx.doi.org/10.1016/S0370-2693(01)00019-3}
  {\path{doi:10.1016/S0370-2693(01)00019-3}}.

\bibitem{Adare:2008sh}
A.~Adare, et~al., {J/psi Production in s(NN)**(1/2) = 200-GeV Cu+Cu
  Collisions}, Phys. Rev. Lett. 101 (2008) 122301.
\newblock \href {http://arxiv.org/abs/0801.0220} {\path{arXiv:0801.0220}},
  \href {http://dx.doi.org/10.1103/PhysRevLett.101.122301}
  {\path{doi:10.1103/PhysRevLett.101.122301}}.

\bibitem{PhysRevD.11.3105}
J.~W. Cronin, H.~J. Frisch, M.~J. Shochet, J.~P. Boymond, P.~A. Pirou\'e, R.~L.
  Sumner, \href{http://link.aps.org/doi/10.1103/PhysRevD.11.3105}{Production of
  hadrons at large transverse momentum at 200, 300, and 400 gev}, Phys. Rev. D
  11 (1975) 3105--3123.
\newblock \href {http://dx.doi.org/10.1103/PhysRevD.11.3105}
  {\path{doi:10.1103/PhysRevD.11.3105}}.
\newline\urlprefix\url{http://link.aps.org/doi/10.1103/PhysRevD.11.3105}

\bibitem{Adam:2015rba}
J.~Adam, et~al., {Inclusive, prompt and non-prompt J/$\psi$ production at
  mid-rapidity in Pb-Pb collisions at $\sqrt{s_{\rm NN}}$ = 2.76 TeV}, JHEP 07
  (2015) 051.
\newblock \href {http://arxiv.org/abs/1504.07151} {\path{arXiv:1504.07151}},
  \href {http://dx.doi.org/10.1007/JHEP07(2015)051}
  {\path{doi:10.1007/JHEP07(2015)051}}.

\bibitem{Aamodt:2010aa}
K.~Aamodt, et~al., {Alignment of the ALICE Inner Tracking System with
  cosmic-ray tracks}, JINST 5 (2010) P03003.
\newblock \href {http://arxiv.org/abs/1001.0502} {\path{arXiv:1001.0502}},
  \href {http://dx.doi.org/10.1088/1748-0221/5/03/P03003}
  {\path{doi:10.1088/1748-0221/5/03/P03003}}.

\bibitem{Alessandro:2006ju}
B.~Alessandro, et~al., {psi-prime production in Pb-Pb collisions at
  158-GeV/nucleon}, Eur. Phys. J. C49 (2007) 559--567.
\newblock \href {http://arxiv.org/abs/nucl-ex/0612013}
  {\path{arXiv:nucl-ex/0612013}}, \href
  {http://dx.doi.org/10.1140/epjc/s10052-006-0153-y}
  {\path{doi:10.1140/epjc/s10052-006-0153-y}}.

\bibitem{Andronic:2009sv}
A.~Andronic, F.~Beutler, P.~Braun-Munzinger, K.~Redlich, J.~Stachel,
  {Statistical hadronization of heavy flavor quarks in elementary collisions:
  Successes and failures}, Phys. Lett. B678 (2009) 350--354.
\newblock \href {http://arxiv.org/abs/0904.1368} {\path{arXiv:0904.1368}},
  \href {http://dx.doi.org/10.1016/j.physletb.2009.06.051}
  {\path{doi:10.1016/j.physletb.2009.06.051}}.

\bibitem{Chen:2013wmr}
B.~Chen, Y.~Liu, K.~Zhou, P.~Zhuang, {$\psi^\prime$ Production and $B$ Decay in
  Heavy Ion Collisions at {LHC}}, Phys. Lett. B726 (2013) 725--728.
\newblock \href {http://arxiv.org/abs/1306.5032} {\path{arXiv:1306.5032}},
  \href {http://dx.doi.org/10.1016/j.physletb.2013.09.036}
  {\path{doi:10.1016/j.physletb.2013.09.036}}.

\bibitem{Khachatryan:2014bva}
V.~Khachatryan, et~al., {Measurement of Prompt $\psi(2S) \to J/\psi$ Yield
  Ratios in Pb-Pb and $p-p$ Collisions at $\sqrt {s_{NN}}=$ 2.76  TeV},
  Phys. Rev. Lett. 113~(26) (2014) 262301.
\newblock \href {http://arxiv.org/abs/1410.1804} {\path{arXiv:1410.1804}},
  \href {http://dx.doi.org/10.1103/PhysRevLett.113.262301}
  {\path{doi:10.1103/PhysRevLett.113.262301}}.

\end{thebibliography}

\end{document}